\documentclass[]{aastex631}

\submitjournal{\apjl, Accepted June 5, 2024~$-$ \href{https://doi.org/10.3847/2041-8213/ad54c5}{doi:10.3847/2041-8213/ad54c5}}

\graphicspath{{./}{figures/}}
\usepackage[version=4]{mhchem}
\usepackage{multirow}
\usepackage{booktabs}
\usepackage{upgreek}
\usepackage{wasysym}
\usepackage{savesym}
\savesymbol{tablenum}
\usepackage{siunitx}
\restoresymbol{SIX}{tablenum}
\usepackage[flushleft]{threeparttable}
\usepackage{comment} %\begin{comment} ... \end{comment}
\begin{document}

\title{Influence of Orbit and Mass Constraints on Reflected Light Characterization of Directly Imaged Rocky Exoplanets}

\correspondingauthor{Arnaud Salvador}
\email{arnaudsalvador@arizona.edu}

\author[0000-0001-8106-6164]{Arnaud Salvador}
\affiliation{Lunar \& Planetary Laboratory, University of Arizona, Tucson, AZ 85721, USA}
\affiliation{Department of Astronomy and Planetary Science, Northern Arizona University, Flagstaff, AZ 86011, USA}
\affiliation{Habitability, Atmospheres, and Biosignatures Laboratory, University of Arizona, Tucson, AZ 85721, USA}
\affiliation{NASA Nexus for Exoplanet System Science Virtual Planetary Laboratory, University of Washington, Box 351580, Seattle, WA 98195, USA}

\author[0000-0002-3196-414X]{Tyler D. Robinson}
\affiliation{Lunar \& Planetary Laboratory, University of Arizona, Tucson, AZ 85721, USA}
\affiliation{Department of Astronomy and Planetary Science, Northern Arizona University, Flagstaff, AZ 86011, USA}
\affiliation{Habitability, Atmospheres, and Biosignatures Laboratory, University of Arizona, Tucson, AZ 85721, USA}
\affiliation{NASA Nexus for Exoplanet System Science Virtual Planetary Laboratory, University of Washington, Box 351580, Seattle, WA 98195, USA}

\author[0000-0002-9843-4354]{Jonathan J. Fortney}
\affiliation{Department of Astronomy and Astrophysics, University of California, Santa Cruz, CA 95064, USA}

\author[0000-0002-5251-2943]{Mark S. Marley}
\affiliation{Lunar \& Planetary Laboratory, University of Arizona, Tucson, AZ 85721, USA}

%%%%%%%%%%%%%%%%%%%%%%%%%%%%%%%%%%%%%%%%%%%%%%%%%%%%%%%%%%%%%%%%%%%%%%%%%%%%%%%%
\begin{abstract}% -- 250 word limit -- %
Survey strategies for upcoming exoplanet direct imaging missions have considered varying assumptions of prior knowledge. Precursor radial velocity surveys could have detected nearby exo-Earths and provided prior orbit and mass constraints. Alternatively, a direct imaging mission performing astrometry could yield constraints on orbit and phase angle of target planets. Understanding the impact of prior mass and orbit information on planetary characterization is crucial for efficiently recognizing habitable exoplanets. To address this question, we use a reflected-light retrieval tool to infer the atmospheric and bulk properties of directly imaged Earth-analogs while considering varying levels of prior information and signal-to-noise ratio (SNR). Because of the strong correlation between the orbit-related parameters and the planetary radius, prior information on the orbital distance and planetary phase angle yield much tighter constraints on the planetary radius: from $R_{\rm{p}}=2.95^{+2.69}_{-1.95}~R_{\oplus}$ without prior knowledge, to $R_{\rm{p}}=1.01^{+0.33}_{-0.19}~R_{\oplus}$ with prior determination of the orbit for $\rm{SNR}=20$ in the visible/near-infrared spectral range, thus allowing size determination from reflected light observations. However, additional knowledge of planet mass does not notably enhance radius ($R_{\rm{p}}=0.98^{+0.17}_{-0.14}~R_{\oplus}$) or atmospheric characterization.
Also, prior knowledge of the mass alone does not yield a tight radius constraint ($R_{\rm{p}}=1.64^{+1.29}_{-0.80}~R_{\oplus}$) nor improves atmospheric composition inference. By contrast, because of its sensitivity to gas column abundance, detecting a Rayleigh scattering slope or bounding Rayleigh opacity helps to refine gas mixing ratio inferences without requiring prior mass knowledge.
Overall, apart from radius determination, increasing the SNR is more beneficial than additional prior observations.

\end{abstract}


%% use the natbib \citep and \citet commands for citations

% =============== Introduction ===============
\section{Introduction} \label{sec:intro}

The discovery and characterization of habitable exoplanets \citep[i.e., planets that can maintain stable liquid water on their surface; e.g.,][]{Kasting1993b} has been defined as one of the leading decadal goals for astronomy, planetary science, and astrobiology \citep{Decadal2021, Decadal_strategy2022}.
The so-called Habitable Worlds Observatory (HWO) mission concept is being developed by NASA to help achieve this profound goal. Importantly, HWO builds on the successes of earlier space-based direct imaging concept missions, such as the Habitable Exoplanet Observatory \citep[HabEx;][]{Gaudi2018, Habex2020} and the Large Ultraviolet/Optical/Infrared Surveyor \citep[LUVOIR;][]{Roberge2018, Luvoir2019}. A coronagraph demonstration as part of the soon-to-launch NASA \textit{Nancy Grace Roman Space Telescope} will also provide key insights into space-based exoplanet high contrast imaging techniques \citep{Akeson2019, Mennesson2020}.

Leveraging recent advances in coronagraph \citep{Trauger2007} or starshade \citep{Harness2021} starlight suppression techniques, HWO is expected to achieve sufficient planet--star flux contrast for exo-Earth direct imaging \citep[below $\sim$10$^{-10}$ in the visible for an Earth-like planet orbiting a Sun-like star at 1 AU; e.g.,][]{Guyon2006, Seager2010}.
Compared to transit spectroscopy, HWO reflected light observations capabilities will probe planets with larger orbital separations from the star and smaller planet-to-star radius ratios \citep[e.g.,][]{Madhusudhan2019}. It will also be sensitive to deeper atmospheric layers \citep[up to the surface or the pressure at which aerosols become optically thick;][]{Morley2015} and even surface conditions \citep[including the detection of oceans' glint;][]{Williams2008, Robinson2010}.
Yet, observations will remain challenging to interpret and retrieve the atmospheric and surface properties encoded in noisy reflected light spectra will not be straightforward.
First and obviously, technical limitations exist \citep[e.g.,][]{Juanola-Parramon2022}; the instrument performance and the observation time restrict the spectral coverage and signal-to-noise ratio needed to sense spectral signatures \citep[e.g.,][]{Feng2018, Konrad2022, Damiano2022, Alei2022, Robinson2023, Susemiehl2023, Latouf2023a, Latouf2023b, Young2024a}. Then, spectral interpretation is challenged by the complexity and our understanding of atmospheric processes \citep[e.g.,][]{Seager2010}, necessarily modeled via simplified parameterizations \citep[e.g.,][]{Fortney2021_chapter, Alei2022b}, and the extremely large range of unconstrained parameters \citep[e.g.,][]{Madhusudhan2018}.
In addition, many degeneracies exist between these parameters and disentangling their respective influence is difficult \citep[e.g.,][]{Benneke2013, Lupu2016, Nayak2017, Welbanks2019, Wang2022_surface}. Indeed, solutions are not unique and many combinations of parameters corresponding to different atmospheric states may satisfactorily reproduce a given observed spectrum.

Some of these observational and theoretical shortcomings are addressed by atmospheric retrieval models, that are used to infer the free/unknown parameters that best reproduce/explain current or future noisy exoplanet observations \citep[e.g.,][]{Madhusudhan2018}.
Recently, and as jointly done for transmission \citep[e.g.,][]{Benneke2012, Greene2016, Lustig-Yaeger2023} and emission spectroscopy \citep[e.g.,][]{VonParis2013b, Konrad2022, Alei2022, Mettler2024}, several retrieval studies have assessed the science returns associated with future direct imaging mission designs targeting planets from gas giants and icy planets \citep{Lupu2016, Nayak2017, Lacy2019, Damiano2020a, Carrion-Gonzalez2020, Damiano2020, Carrion-Gonzalez2021, Damiano2021, Susemiehl2023} to rocky planets \citep{Feng2018, Smith2020, Damiano2022, Robinson2023, Susemiehl2023, Damiano2023, Latouf2023a, Latouf2023b, Young2024a}. In particular, they provide the requirements\,---\,in terms of spectral coverage, resolution, and SNR\,---\,to confidently identify selected atmospheric and surface properties. Thus, atmospheric retrieval modeling can connect our understanding of how well a direct imaging mission can characterize exoplanetary atmospheres to the underlying observational performance delivered by a mission architecture.

Despite these efforts, the most efficient path to characterizing worlds in reflected light is uncertain, especially with regards to the role of prior observations.
Importantly, it has been stated that ``knowledge of a planet’s mass (along with a knowledge of its radius) is essential \lbrack...\rbrack{} to interpret spectroscopic features in its atmosphere'' \citep{NAS_exoplanet_strategy2018}. Yet, this assumption has never been confirmed with retrieval models, nor the values of prior mass knowledge quantitatively assessed.
Extreme precision radial velocity (EPRV) could provide prior detections of exo-Earths, which changes the demands on any discovery survey \citep{Stark2019, Dulz2020, Morgan2021} while yielding orbit and (minimum) mass measurements \citep[e.g.,][]{Lovis2010, Plavchan2015}.
Without EPRV, discovery and characterization requires multiple photometric visits, to first find and select the most promising targets for a follow-up in-depth characterization campaign \citep[e.g., Figure 1.5 from][]{Luvoir2019}. In such a case, the orbit would be determined early to ensure a planet is in its host star's habitable zone.
Alternatively, particularly exciting targets could get marked for spectral observations even before full orbit fits are achieved.
Thus, it is important to understand how atmospheric characterization is influenced by prior knowledge of the orbit and/or mass. 

Here, we investigate how orbit-related and mass prior information will affect our ability to retrieve other atmospheric and planetary properties and help exoplanet characterization. In \autoref{sec:methods}, we describe the retrieval framework and scenarios we considered. Our results showcasing how different levels of prior information at various SNRs can improve exoplanet characterization are presented in \autoref{sec:results}. The main conclusions are summarized in \autoref{sec:conclusion}.

% =============== Methods =============== %
\section{Methods} \label{sec:methods}
We use the \texttt{rfast} atmospheric retrieval suite \citep{Robinson2023} to assess how prior knowledge of the orbit and mass influence the characterization of an Earth-like exoplanet observed with a typical HWO setup.
% general model behavior
The model consists of several interconnected tools. First, a radiative transfer ``forward'' model that generates a synthetic high-resolution spectrum of a planet of known (fiducial) properties. An instrument model then degrades the spectral resolution and adds noise to mimic the ``faux observation'' for a given instrument/telescope design.
A Bayesian sampling tool \citep[\texttt{emcee};][]{Foreman-Mackey2013} then explores the parameter space of all unknown atmospheric and planetary parameters adopted to fit the faux observation.
The set of fit parameters is fed to the forward model that generates the corresponding high-resolution spectrum which is degraded via the instrument model to match the resolution of the simulated data.
The likelihood of the generated spectrum to reproduce the faux observations is computed using the standard definition of chi-squared ($\chi^2$) while the posterior probability informs how the parameter space should be further explored and is computed via Bayes' Theorem. After a burn-in period, the walker positions map the posterior probability distribution. This distribution quantifies which parameter values are most likely to reproduce the simulated observation. Combined with any biases away from the input parameters, this reveals how effectively (or ineffectively) HWO observations could constrain key properties of a directly-imaged world.
Forward and inverse model validations for \texttt{rfast} are presented in \citet{Robinson2023}.

% planet-to-star flux ratio
In high-contrast imaging, the light from the host star that is reflected by the planet is resolved from its bright host as a distinct point source, such that the exoplanet itself is directly imaged.
At a given planet--star--observer phase angle, $\alpha$, the relevant measure is the wavelength-dependent planet-to-star flux ratio:
\begin{equation}
    \frac{F_{\rm p}}{F_{\rm s}} = A_{\rm g} \Phi (\alpha) \left( \frac{R_{\rm p}}{a} \right)^{2}~~,
    \label{eq:Fp_Fs}
\end{equation}
where $A_{\rm g}$ is the geometric albedo, $\Phi$ is the phase function (which depends on the phase angle), $R_{\rm p}$ is the radius of the planet, and $a$ is the orbital distance.
While the planet--star angular separation, the host star apparent magnitude, the exozodiacal dust brightness, and other parameters impact the feasibility of the observation, the planet-to-star flux ratio roughly sets the contrast that must be achieved to image the planet.
To account for the dependence of the reflected light spectrum on the planetary phase, we use \texttt{rfast} 3D mode. The planet is then treated as a homogeneous pixelated globe, where radiative transfer is computed for each pixel following a local plane-parallel assumption. The plane-parallel facets then have different pairs of incidence and emergence angles, depending on the illumination geometry. The total emergent flux from the planet (i.e., from the spatially integrated disk) composed of these many plane-parallel facets is obtained using a Gauss--Chebyshev integration following \citet{Horak1965} \citep[see][for more details]{Robinson2023}.

\begin{figure}
    \centering
    \includegraphics[width=0.8\textwidth, height=1.\textheight, keepaspectratio]{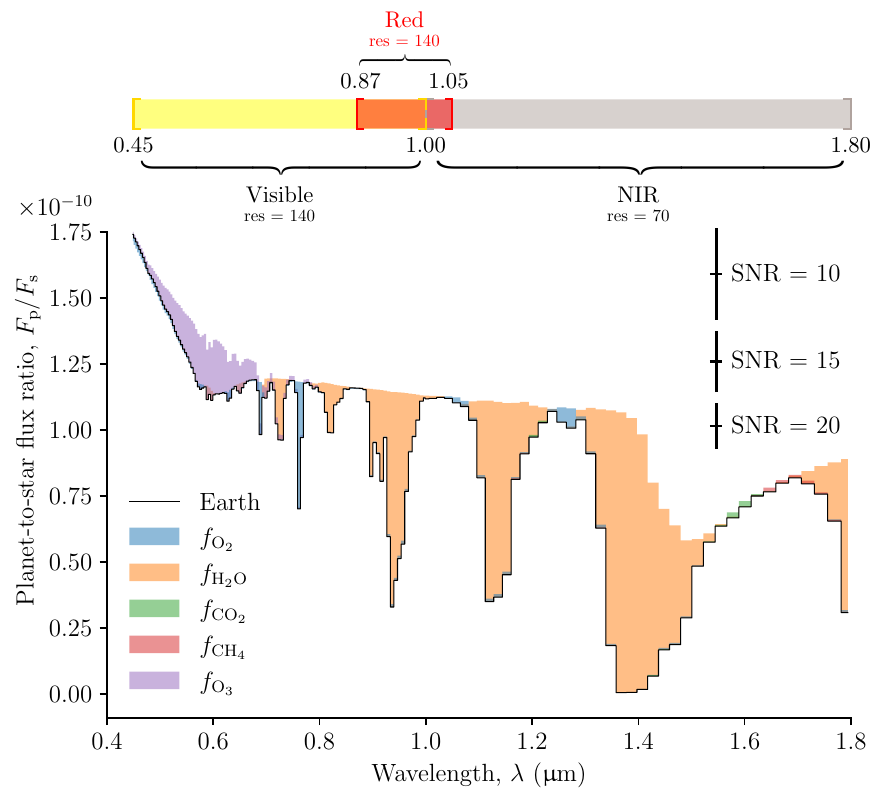}
    \caption{Fiducial model generated (synthetic) reflected light spectrum of an Earth-like planet at quadrature (i.e., at a planetary phase angle $\alpha=90$\textdegree; see \autoref{tab:retrieved_pars} for fiducial input values). The colored areas indicate the spectral impact of the gas species by showing the difference between the baseline spectrum (black line) and the spectrum obtained if they were absent from the atmosphere.
    The top of the figure shows the two wavelength ranges considered and their spectral resolving power: red bandpass ($\rm \lambda = [0.87,~1.05]~\upmu m$, $\rm res = 140$; see \autoref{sec:appendix_spec_cover}) and visible ($\rm \lambda = [0.45,~1]~\upmu m$, $\rm res = 140$) + NIR ($\rm \lambda = [1,~1.80]~\upmu m$, $\rm res = 70$).
    The error bars show the nonrandomized, wavelength-independent noise corresponding to specific signal-to-noise ratios (SNR) at the reference wavelengths $\rm \lambda = 0.45~\upmu m$ for the vNIR range, and $\rm \lambda = 0.88~\upmu m$ for the red bandpass.
    }
    \label{fig:fiducial_spectrum}
\end{figure}

% faux observations Earth-like fiducial parameters
Because our focus is on the characterization of Earth-like habitable planets with future direct imaging mission concepts, we simulated the reflected light spectrum of the best characterized habitable planet: Earth. Our fiducial parameter values used to generate these synthetic observations are given in \autoref{tab:retrieved_pars}, where the atmosphere is made primarily of nitrogen (78\%) and oxygen (21\%), with water vapor, carbon dioxide, methane, ozone, and argon as additional trace species. The Earth analog simulated spectrum is passed to a simple instrument model \citep{Robinson2016} to mimic observations relevant to HWO (\autoref{fig:fiducial_spectrum}).

% retrieved parameters
The \texttt{rfast} inverse tool has been designed for the rapid exploration of many atmospheric and planetary parameters without being computationally prohibitive.
Here we retrieved on 17 parameters (listed in \autoref{tab:retrieved_pars} with the input value and prior range considered) whose contributions are encoded in the spectrum. The surface conditions are described via the surface pressure ($p_{\rm surf}$), the isothermal atmospheric temperature ($T$), and the gray surface albedo ($A_{\rm surf}$). The atmospheric composition is set by the gas abundances of \ce{N2}, \ce{O2}, \ce{H2O}, \ce{CO2}, \ce{CH4}, and \ce{O3} (expressed in terms of atmospheric volume mixing ratios, $f$), where argon fulfills the rest of the atmosphere when the sum of their volume mixing ratios is lower than one. The properties of clouds are captured by their top pressure ($p_{\rm c}$), thickness ($\Delta p_{\rm c}$), optical depth ($\tau_{\rm c}$), and the cloudiness fraction ($f_{\rm c}$). The planetary bulk parameters are the planet radius ($R_{\rm p}$) and mass ($M_{\rm p}$). Finally, the orbit-related parameters are the orbital distance ($a$) and the planet--star--observer phase angle ($\alpha$).
\texttt{emcee} was run with 15 Markov Chain Monte Carlo (MCMC) chains (walkers) per parameter (for a total of 255 walkers). For each case, we let \texttt{emcee} explore the parameter space for 100,000 steps. Sensitivity studies conducted while developing \texttt{rfast} demonstrated that the chains have then fully converged, and this for all retrieved parameters \citep[][]{Robinson2023}. Drawing the posterior distributions from the last 5000 steps gives a sufficiently large sample to be statistically representative without being computationally prohibitive. It also maintains consistency with \citet{Feng2018, Robinson2023}, and ensures that the walkers have properly explored the parameter space and forgotten their initial position.

% prior information
To assess the influence of orbital and mass prior information on exoplanet characterization, we consider three different retrieval scenarios associated with different levels of prior constraints: (1) ``no prior constraint'', i.e., none of the retrieved parameters is known a priori; (2) ``known orbit'', where a prior determination of the orbit of the planet would give constraints on the orbital distance ($a$) and planetary phase angle ($\alpha$) at the time of observation such that these are constrained to 10\% of Earth's value \citep[consistent with HabEx/LUVOIR/HWO astrometry target performance; e.g.,][]{Guyon2013, Horning2019, Luvoir2019, Habex2020}; (3) ``known orbit and mass'', where both the orbit-related parameters ($a$, $\alpha$) and the mass of the planet ($M_{\rm p}$) have been previously informed and are all constrained to 10\% of Earth's value \citep[consistent with predicted precision radial velocity semi-amplitude uncertainties of 10\%;][]{Plavchan2015}. An additional ``mass constrained'' case, where only the mass is constrained by prior information, is also shown for a single value of signal-to-noise ratio (SNR = 20) and our baseline spectral coverage (visible/near-infrared) to isolate the influence of mass prior determination.
Priors are taken as Gaussian constraints. The ``known orbit'' case models a scenario where the observatory has used earlier observations to help astrometrically constrain the orbit while the ``known orbit and mass'' case mimics a scenario where precision radial velocity or astrometric data has obtained an earlier mass determination.
Note that none of the scenarios assume any specific mass--radius relationship. Adopting such a relationship as a prior could help to further limit the allowable phase space.

% observation setup/instrument design -- spectral coverages, R, SNR
Our instrument model reproduces the observational capabilities of a typical HWO setup operating in the ``vNIR'' spectral range: an optimistic combination of ``visible'' ($\rm \lambda = [0.45,~1]~\upmu m$, with a spectral resolving power $\rm res = \lambda/\delta \lambda = 140$) and near-infrared (NIR; $\rm \lambda = [1,~1.80]~\upmu m$, $\rm res = 70$) wavelengths (as could be obtained with multiple starshades and/or coronagraph pointings) (\autoref{fig:fiducial_spectrum}). While it seems reasonable to expect observations spanning the full vNIR bandpass to maximize the yield at the stage of detailed exoplanet characterization, we also tested how our results would apply to observations conducted in a ``red'' ($\rm \lambda = [0.87,~1.05]~\upmu m$, $\rm res = 140$) coronagraph-restricted bandpass including the prominent 0.94 $\rm \upmu m$ \ce{H2O} absorption feature (\autoref{fig:fiducial_spectrum}).

For the vNIR baseline spectral coverage, we consider a constant, nonrandomized, gray (i.e., wavelength-independent) noise for three different signal-to-noise ratio (SNR) values (10, 15, and 20) to account for different data quality without being restricted to specific observing parameters such as telescope diameter or target distance (compared to integration times\citealp[; see][]{Feng2018}). Note that for a constant noise, the SNR goes down within spectral features. The SNR values of 10, 15, or 20 refer to the SNR set at $\rm \lambda = 0.45~\upmu m$, which is then propagated across the entire spectral coverage while maintaining a constant error bar size.
For the pessimistic red bandpass test, we consider SNR = 10 only (specified at 0.88 $\rm \upmu m$).
This nonrandomized (i.e., the error bars are simply centered on truth/noise-free values), wavelength-independent noise model follows the approach described in \citet{Feng2018} for their initial validation. \citet{Feng2018} demonstrated that for a statistical sampling of retrievals, a more realistic, randomized noise would yield similar inference results. Furthermore, a constant noise aligns with the predicted performance of the HabEx and LUVOIR missions, assuming equal exposure times across their UV, optical, and NIR spectral bands \citep[][]{Luvoir2019, Habex2020}. The gray noise approach also enables easier reproducibility, as the wavelength-dependent noise need not be specified.
Note that we will specifically address the influence of the data quality (both in terms of spectral coverage and SNR) on atmospheric retrievals and in determining the best observing strategy in another dedicated study.

Considering the different levels of prior information and the variations in SNR and spectral coverage, we ran a total of 13 retrieval scenarios and discuss how these scenarios allow to infer the properties of an observed Earth analog in the next section.

\begin{table}
\centering
\caption{Retrieved parameters, corresponding description, Earth-based fiducial input value, and prior ranges}
\label{tab:retrieved_pars}
\resizebox{1\textwidth}{!}{%
\begin{tabular}{@{}lllcc@{}}
\toprule %-----------------------------------------------------------------------------------------------------------------------------------
Parameter              & Description                         & Input                 & Flat prior                   & Gaussian prior$^\ddag$ \\
\midrule %-----------------------------------------------------------------------------------------------------------------------------------
\multicolumn{5}{c}{\textit{Surface conditions}}                                                                                            \\
log $p_{\rm surf}$     & Surface pressure (Pa)               & log($10^5$)           & [0,~8]             & ---                       \\
$T$                    & Atmospheric temperature$^*$ (K)     & 255                   & [100,~1000]             & ---                       \\
log $A_{\rm surf}$     & Surface albedo                      & log(0.05)             & [$-2$,~0]               & ---                       \\
\midrule %-----------------------------------------------------------------------------------------------------------------------------------
\multicolumn{5}{c}{\textit{Gas abundances$^\dag$}}                                                                                         \\
log $f_{\rm\ce{N2}}$   & Molecular nitrogen mixing ratio     & log(0.78)             & [$-10$,~0]         & ---                       \\
log $f_{\rm\ce{O2}}$   & Molecular oxygen mixing ratio       & log(0.21)             & [$-10$,~0]         & ---                       \\
log $f_{\rm\ce{H2O}}$  & Water vapor mixing ratio            & log($3\times10^{-3}$) & [$-10$,~0]         & ---                       \\
log $f_{\rm\ce{CO2}}$  & Carbon dioxide mixing ratio         & log($4\times10^{-4}$) & [$-10$,~0]         & ---                       \\
log $f_{\rm\ce{CH4}}$  & Methane mixing ration               & log($2\times10^{-6}$) & [$-10$,~0]         & ---                       \\
log $f_{\rm\ce{O3}}$   & Ozone mixing ratio                  & log($7\times10^{-7}$) & [$-10$,~$-2$] & ---                       \\
\midrule %-----------------------------------------------------------------------------------------------------------------------------------
\multicolumn{5}{c}{\textit{Cloud parameters}}                                                                                              \\
log $p_{\rm c}$        & Cloud-top pressure (Pa)             & log($6\times10^{4}$)  & [0,~8]             & ---                       \\
log $\Delta p_{\rm c}$ & Cloud thickness (Pa)                & log($10^4$)           & [0,~8]             & ---                       \\
log $\tau_{\rm c}$     & Cloud optical depth                 & log(10)               & [$-3$,~$3$]   & ---                       \\
log $f_{\rm c}$        & Cloudiness fraction                 & log(0.5)              & [$-3$,~0]          & ---                       \\
\midrule %-----------------------------------------------------------------------------------------------------------------------------------
\multicolumn{5}{c}{\textit{Planetary bulk parameters}}                                                                                     \\
log $R_{\rm p}$        & Planet radius ($R_{\oplus}$)        & log(1)                & [$-1$,~$1$]               & ---                       \\
log $M_{\rm p}$        & Planet mass ($M_{\oplus}$)          & log(1)                & [$-1$,~$2$]              & \{1, 0.1\}                \\
\midrule %-----------------------------------------------------------------------------------------------------------------------------------
\multicolumn{5}{c}{\textit{Orbital parameters}}                                                                                            \\
$a$                    & Planetary orbital distance (AU)     & 1                     & [0.1,~10]               & \{1, 0.1\}                \\
$\alpha$               & Planetary phase angle (\textdegree) & 90                    & [0,~180]                & \{90, 9\}                 \\
\bottomrule %--------------------------------------------------------------------------------------------------------------------------------
\multicolumn{5}{l}{$^*$Isothermal atmosphere temperature.}\\
\multicolumn{5}{l}{$^\dag$The remaining atmosphere is back-filled with argon.}\\
\multicolumn{5}{p{13cm}}{$^\ddag$Parameters constrained to 10\% of Earth's value when considering prior observations. The prior distribution is then a Gaussian $\{ \mu, \sigma \}$ of mean $\mu$ centered on Earth's value, with a standard deviation $\sigma$ of 10\% the truth value.
}\\
\end{tabular}%
}
\end{table}

% =============== Results ===============
\section{Results and Discussion} \label{sec:results}
Material below details the value of different types of prior information. First we explore the impact of priors on determinations of the planetary radius. Next we highlight how the inference of key parameters is influenced by prior information versus increasing the SNR.

\subsection{Planet Radius Determination}
\begin{figure}
    \centering
    \includegraphics[width=0.8\textwidth, height=1.\textheight, keepaspectratio]{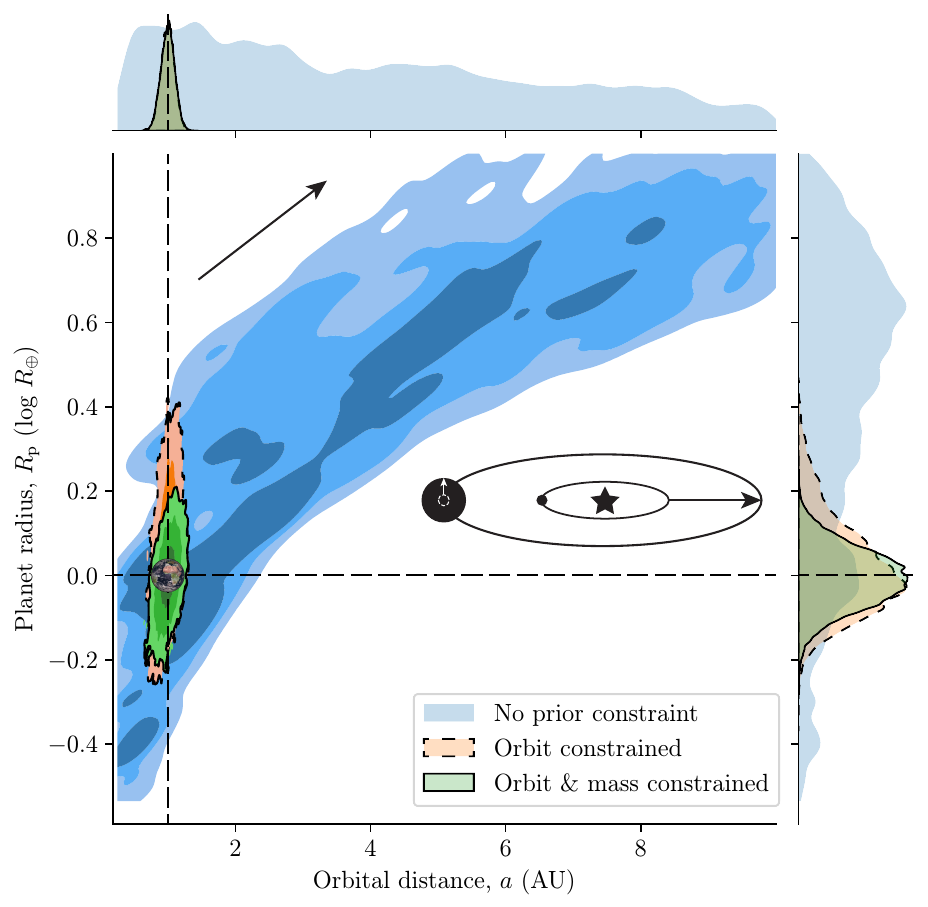}
    \caption{Orbital distance and planet radius joint/bivariate (center) and marginal/univariate (top and right side) posterior distributions for different scenarios of prior knowledge and for the vNIR and $\rm SNR = 20$.
    The contours of the 2D posterior distributions denote the 1, 2, and 3 sigma levels, encompassing 68\%, 95\%, and 99.7\% of the observed values, respectively (note that the relevant 1, 2, and 3 sigma levels for a 2D distribution of samples are 39.3\%, 86.5\%, 98.9\% of the volume, and correspond to the 68\%, 95\%, and 99.7\% for 1D distributions).
    Earth-like input values are depicted with dashed horizontal and vertical lines. The small image of Earth indicates its position in the diagram.
    The sketch illustrates the degenerate relationship between increasing orbital distance and increasing planet radius, in yielding equivalent amount of reflected light. A larger planet orbiting further away from its host star may essentially reflect the same amount of light than a smaller planet but closer from the star. As a result, if the orbit-related parameters are known or constrained, the range of planet radius yielding the same amount of reflected light is necessarily restricted.
    Retrieval results are given in \autoref{tab:ret_results_vNIR_SNR20}.
    }
    \label{fig:orbit_radius}
\end{figure}

% orbit--radius correlation
\autoref{fig:orbit_radius} illustrates the degenerate relationship between increasing orbital distance and increasing planetary radius, in yielding an equivalent amount of reflected light. Indeed, because the flux received (and reflected) by a planet increases with decreasing the orbital distance and with increasing the planet radius, a larger planet orbiting further away from its host star may essentially reflect the same amount of light than a smaller planet but closer from the star. This correlation follows directly from the ratio of planetary size to orbital distance in \autoref{eq:Fp_Fs}. More subtly, \citet{Nayak2017} showed that a similar degenerate relationship exists between the planet phase and radius. Increasing the planet radius may reflect essentially the same amount of light and compensates for a decreasing planet phase (increasing phase angle).
Because the geometric albedo ($A_{\rm g}$) and the phase function ($\Phi$) are physically bound via optical scattering and absorption, measuring the planet-to-star flux ratio while knowing the planetary phase angle and the orbital distance would then ultimately enable inferences of the planet radius (\autoref{eq:Fp_Fs}).

\begin{figure}
    \centering
    \includegraphics[width=0.8\textwidth, height=1.\textheight, keepaspectratio]{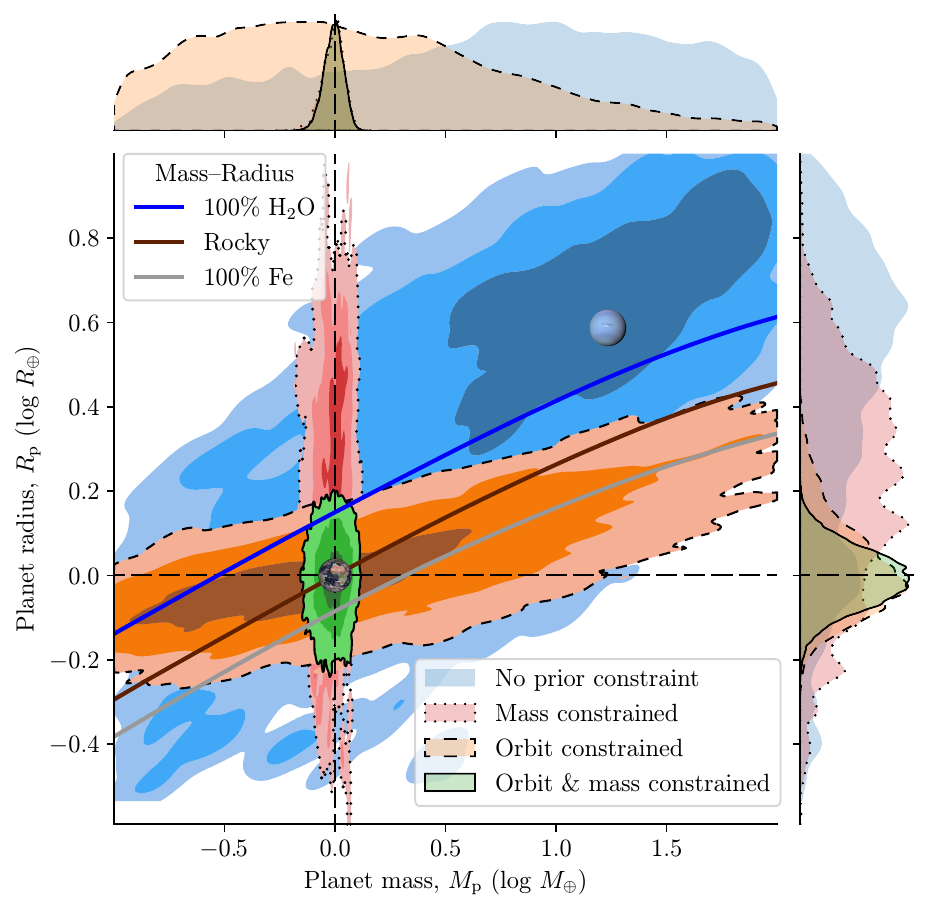}
    \caption{Mass--radius diagram (center) and marginal univariate posterior distributions (top and right side) for the different scenarios of prior knowledge and for the vNIR and $\rm SNR = 20$ case.
    The contours of the 2D posterior distributions denote the 1, 2, and 3 sigma levels, encompassing 68\%, 95\%, and 99.7\% of the observed values, respectively.
    Earth-like input values are depicted with dashed horizontal and vertical lines. The small images of Earth and Neptune indicate their positions in the diagram.
    For comparison, mass--radius relationships representative of different bulk compositions and interior structures are shown (pure \ce{H2O} assuming 1 mbar surface pressure level at 300 K, Earth-like rocky: 32.5\% \ce{Fe} + 67.5\% \ce{MgSiO3}, and pure iron; from \citealt{Zeng2019}\footnote{\url{https://lweb.cfa.harvard.edu/~lzeng/planetmodels.html##mrrelation}}). Retrieval results are given in \autoref{tab:ret_results_vNIR_SNR20}.
    }
    \label{fig:mass_radius}
\end{figure}

A prior determination of the orbit would give constraints on the orbital distance and planetary phase angle at the time of observation.
Thus, in such a case, the planet radius required to match the amount of reflected light becomes tightly constrained (orange area in \autoref{fig:orbit_radius}).
\autoref{fig:mass_radius} shows posterior constrains on the planetary radius and type as a function of the prior information available for vNIR at SNR = 20 and clearly illustrates this behavior.
Without any prior knowledge on the retrieved parameters (blue areas in \autoref{fig:mass_radius}), and without any mass--radius relationship assumed, the full prior range for the planet radius is explored ($\log R_{\rm p}/R_{\oplus} = 0.47^{+0.28}_{-0.47}$, i.e., the radius is confidently constrained to be within 0.99 and 5.63 $R_{\oplus}$ for the 1-sigma or 68\% credible interval), meaning that any radius may satisfactorily reproduce the observed spectrum. As seen from the marginal posterior distribution that weakly peaks above Neptune-sized planets ($R_{\neptune}=3.865~ R_{\oplus} = 0.587~\log R_{\oplus}$), large planets could be mistakenly favored over Earth-sized planets.

Because of the orbit--radius correlation, prior knowledge of the orbit significantly restricts the range of likely radius, i.e., reproducing the observations, yielding accurate radius determination. The radius is confidently constrained to be between 0.81 and 1.34 Earth radius ($\log R_{\rm p}/R_{\oplus} = 0.00^{+0.12}_{-0.09}$ for the 1-sigma or 68\% credible interval). The planet is then properly and without ambiguity identified as an Earth-sized planet.
Prior knowledge of the orbit also yields a tight radius characterization in all retrieval scenarios conducted, including for different SNR values in the vNIR bandpass (\autoref{fig:post_dist_key_parameters}). Even for a reduced red bandpass ($\rm \lambda = [0.87,~1.05]~\upmu m$, with a spectral resolving power res = 140; see \autoref{fig:fiducial_spectrum}) and a $\rm SNR = 10$ (specified at $\rm \lambda = 0.88~\upmu m$), the planet radius is confidently constrained to be between 0.63 and 1.73 Earth radius when the orbit is already determined ($\log R_{\rm p}/R_{\oplus} = -0.02^{+0.25}_{-0.19}$ for the 68\% confidence interval, corresponding to $R_{\rm p} = 0.96^{+0.77}_{-0.33}~ R_{\oplus}$), while it is unconstrained ($R_{\rm p} = 2.77^{+3.37}_{-1.91}~ R_{\oplus}$) without prior knowledge (see \autoref{fig:mass_radius_Red_SNR10} and \autoref{tab:ret_results_Red_SNR10} in \autoref{sec:appendix_spec_cover}). Importantly, it shows that these results are not strongly dependent on the wavelength coverage or SNR considered.
Note that characterizing the radius this way would reduce important degeneracies involving it and thus help constraining other parameters \citep{Carrion-Gonzalez2020}.

Because of the already tight constraints on the radius, the mass prior, added to the orbit-related priors, does not offer substantial improvements on the planetary radius constraint, neither for observations in the vNIR ($\log R_{\rm p}/R_{\oplus} = -0.01^{+0.07}_{-0.07}$, corresponding to $R_{\rm p} = 0.98^{+0.17}_{-0.14}~ R_{\oplus}$; see \autoref{fig:mass_radius}) nor in the red bandpass ($\log R_{\rm p}/R_{\oplus} = -0.03^{+0.23}_{-0.16}$, corresponding to $R_{\rm p} = 0.93^{+0.67}_{-0.29}~ R_{\oplus}$; see \autoref{fig:mass_radius_Red_SNR10}). However, such mass information could prove valuable when attempting to holistically understand the world, as the added mass information helps to constrain bulk properties like density. Here, with both orbit-related and mass priors, the planet is unequivocally identified as rocky.
However, results where only the mass is constrained by prior information (red areas in \autoref{fig:mass_radius} and \autoref{fig:post_dist_key_parameters}, and \autoref{tab:ret_results_vNIR_SNR20}) indicate that knowing only the planet mass would not be sufficient to firmly constrain the planet radius ($\log R_{\rm p}/R_{\oplus} = 0.22^{+0.25}_{-0.29}$, corresponding to a planet of 0.84 to 2.94 Earth radii at the 68\% confidence interval) due to the weaker $M_{\rm p}-R_{\rm p}$ correlation, but rules out Neptune-sized planets (3.865\,$R_{\oplus}$) at the 68\% credible interval.

\subsection{Atmospheric Properties and Surface Conditions Inference}
\begin{figure}
    \centering
    \includegraphics[width=1\textwidth, height=0.9\textheight, keepaspectratio]{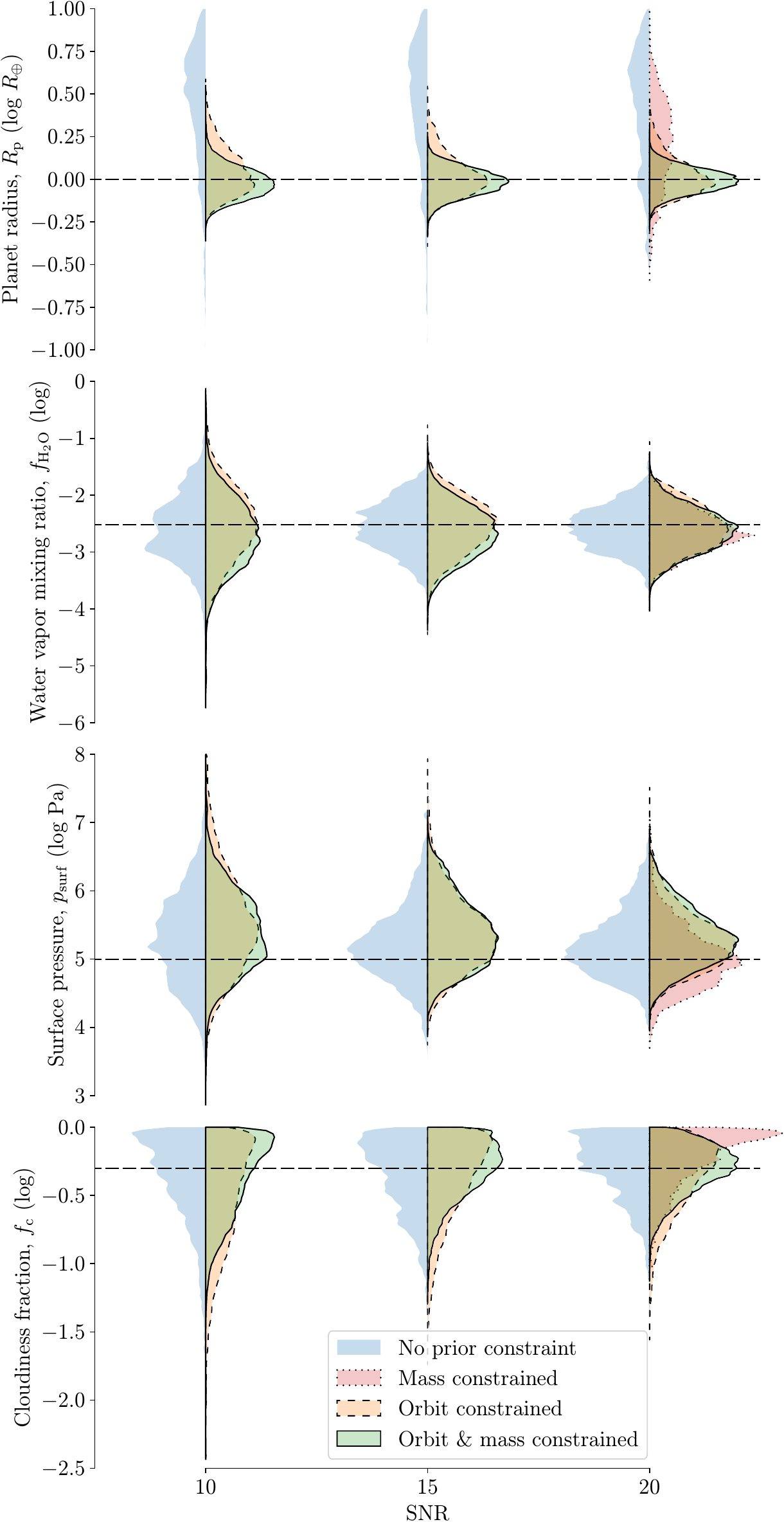}
    \caption{Planet radius ($R_{\rm p}$), water vapor mixing ratio ($f_{\rm H_2O}$), surface pressure ($p_{\rm surf}$), and cloud fraction ($f_{\rm c}$) posterior distributions for the different levels of color-coded prior knowledge, at different SNR values (10, 15, 20) in the vNIR. The ``mass constrained'' case (i.e., where only the mass is constrained; in red) is shown for $\rm SNR=20$ only. 
    Earth-like input values are depicted with dashed horizontal lines.
    The density has been normalized across all violins such that each have the same area.
    Apart from the planet radius inference, which is significantly improved by prior knowledge, increasing the SNR is more beneficial for atmospheric characterization; as illustrated by the shrinking tails of the posterior distributions and their narrowing towards the fiducial values with increasing SNR (see also \autoref{tab:ret_results_vNIR_SNR20}, \autoref{tab:ret_results_vNIR_SNR15}, and \autoref{tab:ret_results_vNIR_SNR10} for the retrieval results).
    }
    \label{fig:post_dist_key_parameters}
\end{figure}

The influence of prior observations on the inference of key parameters ($R_{\rm p}$, $f_{\rm H_2O}$, $p_{\rm surf}$, $f_{\rm c}$) is shown in \autoref{fig:post_dist_key_parameters} for the vNIR, while retrieval results for the full suite of parameters at SNR = 20, 15, and 10 are given in \autoref{sec:appendix_retrieval_results}, in \autoref{tab:ret_results_vNIR_SNR20}, \autoref{tab:ret_results_vNIR_SNR15}, and \autoref{tab:ret_results_vNIR_SNR10}, respectively.
The corresponding posterior distributions obtained for different SNR are also shown to assess the relative value of adding more SNR (i.e., integrating longer) against adding more prior observations. For additional details, univariate and bivariate posterior distributions of the full suite of retrieved parameters for observations in the vNIR at SNR = 20, and for the different scenarios of prior knowledge, are provided as a corner plot in \autoref{fig:corner_vNIR_SNR20} of \autoref{sec:appendix_retrieval_results}.

% Influence of priors
Except for the planet radius and regardless of the SNR value, prior knowledge of the orbit-related parameters and/or of the mass does not significantly improve retrieval accuracy and atmospheric/exoplanet characterization (\autoref{fig:post_dist_key_parameters}). This contrasts with the idea that ``knowledge of a planet’s mass (along with a knowledge of its radius) is essential \lbrack...\rbrack{} to interpret spectroscopic features in its atmosphere'' \citep{NAS_exoplanet_strategy2018}. Only the cloudiness fraction ($f_{\rm c}$) inference is noticeably refined by prior determination of the mass, and this only when the orbit is already informed: adding the mass prior to the orbit-related knowledge shrinks the tail of the posterior and shifts its peak towards the fiducial value, which allows to constrain $f_{\rm c}$ for $\rm SNR = 15$ (green compared to orange area of the middle and bottom panel of \autoref{fig:post_dist_key_parameters} and \autoref{tab:ret_results_vNIR_SNR15}).

On the other hand, the top pressure of the cloud deck ($p_{\rm c}$) inference is only marginally refined when adding the mass prior to the already informed orbit at $\rm SNR = 10$ (\autoref{tab:ret_results_vNIR_SNR10}). Its posterior distribution peaks around the fiducial value for all scenarios and $p_{\rm c}$ only becomes ``constrained'' because the uncertainties are slightly reduced and fall below the 1 log-unit threshold of our inference accuracy classification when adding the mass prior (\autoref{sec:appendix_retrieval_results}).

To understand the role of the mass prior, recall that planetary mass directly impacts surface gravity, and surface gravity is inversely proportional to the column abundance of molecules above a given pressure. As absorption features are roughly sensitive to the product of the opacity and the column abundance (which sets the atmospheric transmissivity), one might expect that gas mixing ratio constraints would correlate strongly with gravity and mass, such that mass prior information would translate into much-improved atmospheric composition constraints. However, Rayleigh scattering optical depth is directly sensitive to the bulk column abundance of the atmosphere, thereby helping to prevent a degeneracy between planetary mass and atmospheric composition.
Observations detecting a Rayleigh scattering slope in the visible thus inform the bulk column abundance and allow constraints on gas mixing ratios without requiring prior mass knowledge.
Improving observational quality for spectral features by increasing the SNR thus yields better constraints on the mixing ratio whereas prior information on the mass does not.

Without measuring the Rayleigh scattering slope\,---\,such as for observations in the red bandpass\,---\,constraints on atmospheric composition worsen (e.g., \autoref{tab:ret_results_Red_SNR10}) because of the degeneracy between planetary mass and atmospheric column abundance. Note that detecting a Rayleigh scattering slope is not critically required to constrain gas mixing ratios absent mass information, as even upper limits on Rayleigh scattering opacity can provide corresponding upper limits on the bulk gas column abundance. 

Thus, future direct imaging missions should strive to provide observations at the shortest feasible wavelengths, especially if the likelihood of prior mass information is uncertain or unlikely. This with the caveat that atmospheric hazes could potentially disrupt the Rayleigh signature. Future work should investigate how assumptions of background gases and hazes, and their Rayleigh scattering, impact the atmospheric characterization-related utility of mass priors.

Alternatively, and while it is always constrained otherwise for $\rm SNR = 20$ even without prior constraints (\autoref{tab:ret_results_vNIR_SNR20}), prior knowledge of the orbit only worsens the inference of the cloudiness fraction. In this scenario, the extension of the posterior distribution's tail to lower values (orange area of the bottom right panel of \autoref{fig:post_dist_key_parameters}) results in a 68\% confidence interval slightly wider than 0.5 log-units, surpassing our ``constrained'' threshold (see \autoref{sec:appendix_retrieval_results}).

% Influence of increasing the SNR
Increasing the SNR systematically improves atmospheric characterization, as discussed in \citet{Feng2018}. 
This is illustrated by the fact that the posterior distributions narrow towards the fiducial values and their tails shrink with increasing SNR (see also \autoref{tab:ret_results_vNIR_SNR20}, \autoref{tab:ret_results_vNIR_SNR15}, and \autoref{tab:ret_results_vNIR_SNR10} and the width of the 68\% credible intervals).
Here, a SNR of at least 15 is required to constrain the water vapor mixing ratio ($f_{\rm H_2O}$) and the surface pressure ($p_{\rm surf}$) becomes constrained from SNR = 20 (\autoref{fig:post_dist_key_parameters} and \autoref{sec:appendix_retrieval_results}).

Overall, apart from prior knowledge of the orbit that allows for accurate radius determination from reflected-light spectroscopy, these results indicate that orbit and/or mass priors do not significantly improve retrieval efficiency for Earth-like worlds, and that increasing the SNR is much more beneficial to atmospheric characterization.

Note that atmospheres significantly differing from Earth-like worlds, such as those of completely haze/cloud-covered terrestrials or gaseous worlds without a solid surface, might have different behaviors for which the applicability of these results should be further tested.
In addition, our constraints may change as the architecture for HWO is refined and as we attain a more comprehensive understanding of the noise characteristics. A yield modeling tool \citep[e.g.,][]{Delacroix2016, Stark2019, Morgan2021} could then be upgraded to assess the observing strategies that balance orbit determination against atmospheric characterization in scenarios where mass priors are provided or not.

% =============== Conclusion =============== %
\section{Conclusion} \label{sec:conclusion}
The detection and characterization of Earth-like planets orbiting within the habitable zone of Sun-like stars has been set as a major goal for the next decade of planetary science. With it, future missions such as the Habitable Worlds Observatory are being designed.
Using \texttt{rfast} in a typical Habitable Worlds Observatory setup, we investigated how prior knowledge of the planet mass and orbit-related parameters will affect the characterization of Earth analogs observed in reflected light.

If the orbit of the planet is known a priori, we found that the planet radius can be accurately retrieved from reflected light observations, regardless of the SNR or spectral coverage considered here. Combined with prior determination of the planet mass derived from radial velocity measurements, these indirect radius constraints would allow for the firm determination of the planet type and density, thus providing key information to assess its composition, interior structure, and contextualizing the observations in the frame of coupled interior-atmosphere evolution leading to the observed atmospheric state. 

While remaining essential to infer fundamental properties, prior knowledge of the mass alone or combined with orbit-related priors does not significantly improve atmospheric characterization nor surface conditions inference.
Because Rayleigh scattering optical depth is directly sensitive to the bulk column abundance of the atmosphere, detecting a Rayleigh scattering slope or bounding Rayleigh opacity helps to refine gas mixing ratio inferences without requiring prior mass knowledge. Future direct imaging missions should thus strive to provide observations at the shortest feasible wavelengths, especially if the likelihood of prior mass information is uncertain or unlikely.
Overall, and apart from radius determination, increasing the SNR of given observations is more beneficial than adding more prior observations to characterize an exoplanet and infer its atmospheric composition.

% ============ Acknowledgments ============= %
%\begin{acknowledgments}
\section*{Acknowledgments}
AS and TDR gratefully acknowledge support from NASA's Habitable Worlds Program (No.~80NSSC20K0226) and from NASA's Nexus for Exoplanet System Science Virtual Planetary Laboratory (No.~80NSSC18K0829). TDR also acknowledges support the Cottrell Scholar Program administered by the Research Corporation for Science Advancement.
The authors thank an anonymous reviewer for constructive comments that increased the accessibility of the manuscript.
%\end{acknowledgments}

\vspace{5mm}
\software{
\texttt{Astropy} \citep{astropy:2013, astropy:2018, astropy:2022},
\texttt{corner} \citep{Foreman-Mackey2016},
\texttt{emcee} \citep{Foreman-Mackey2013},
\texttt{matplotlib} \citep{matplotlib},
\texttt{NumPy} \citep{numpy},
\texttt{pandas} \citep{pandas},
\texttt{rfast} \citep{Robinson2023},
\texttt{seaborn} \citep{seaborn}
}

% =============== Appendix =============== %
%\clearpage
\appendix

\section{Retrieval results for vNIR at different SNR} \label{sec:appendix_retrieval_results}

To describe atmospheric inference accuracy, we use a detection strength classification similar to \citet{Feng2018}. Retrieved parameters can be 
\begin{itemize}
    \item[--] Tightly Constrained (``TC'' -- for lin-scale prior ranges only): the width of the 68\% confidence interval (i.e., the range of values within -1 and +1 standard deviation, sigma, from the mean of a Gaussian distribution) is smaller than $\pm10\%$ of the input value;
    \item[--] Constrained (``C''): for log-scale prior ranges, the width of the 68\% confidence interval is contained within 1 log-unit (i.e., within one order of magnitude) for prior ranges spanning at least 5 log-units, and within 0.5 log-units otherwise.
    For lin-scale prior ranges, the width of the 68\% confidence interval is contained within 10\% of the full prior range width (\autoref{tab:retrieved_pars}). This is typically the case for posterior distributions with a marked peak without tails toward extreme values.
    \item[--] Non Constrained (``NC''): when none of the above condition is met. This is typically the case when the posterior distribution is flat across the entire (or nearly entire) prior range, or has a substantial tail towards extreme values.
\end{itemize}
This classification is only relevant to parameters of flat priors. It is not applicable (``NA'') to parameters known a priori and artificially constrained using a Gaussian-shaped, informative prior centered on the truth value (see ``Gaussian prior'' column in \autoref{tab:retrieved_pars}).

\autoref{tab:ret_results_vNIR_SNR20}, \autoref{tab:ret_results_vNIR_SNR15}, and \autoref{tab:ret_results_vNIR_SNR10} provide the retrieval results obtained for the different prior observations available for observations conducted in the vNIR at SNR of 20, 15, and 10, respectively, along with the detection strength associated with each retrieved parameter.

\begin{figure}
    \centering
    \includegraphics[width=1\textwidth, height=1\textheight, keepaspectratio]{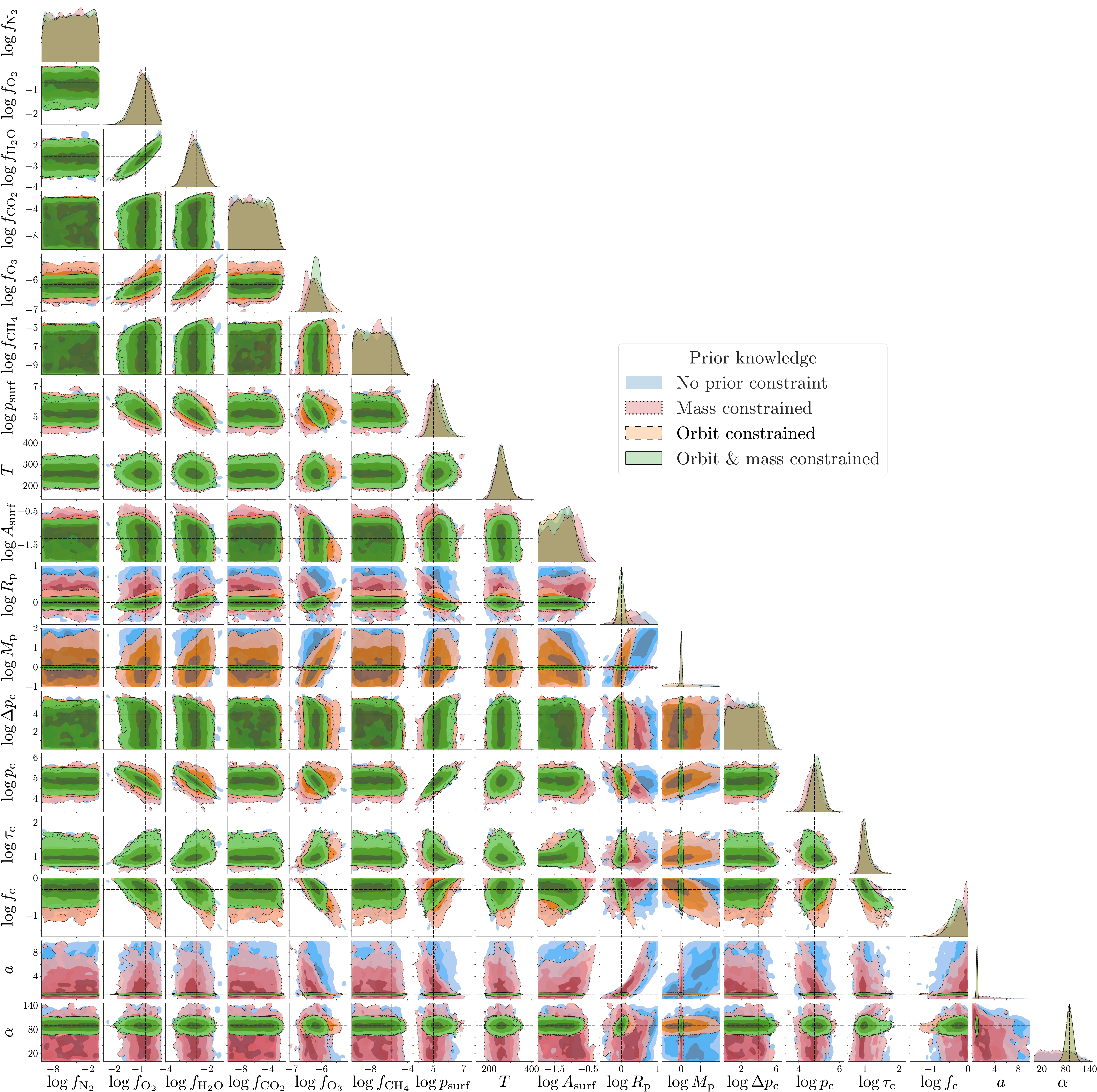}
    \caption{Corner plot illustrating the marginal univariate (along the diagonal) and bivariate (off-diagonal) posterior distributions for all retrieved parameters, considering various scenarios of prior knowledge for observations conducted in the vNIR at SNR = 20 (see \autoref{tab:ret_results_vNIR_SNR20} for quantitative estimates). 
    The contours of the 2D posterior distributions denote the 1, 2, and 3 sigma levels, encompassing 68\%, 95\%, and 99.7\% of the observed values, respectively.
    Earth-like input values are depicted with dashed horizontal and vertical lines.
    }
    \label{fig:corner_vNIR_SNR20}
\end{figure}

\autoref{fig:corner_vNIR_SNR20} illustrates the marginal univariate (along the diagonal) and joint bivariate (off-diagonal) posterior distributions of all retrieved parameters for various scenarios of prior knowledge, based on observations in the vNIR at SNR = 20.

\begin{table}
\centering
\resizebox{1\textwidth}{!}{%
\begin{threeparttable}
\caption{Retrieval results comparison and associated level of constraint for the different cases of prior knowledge considered for vNIR and SNR = 20}
\label{tab:ret_results_vNIR_SNR20}
\begin{tabular}{lS|S@{\hspace{-0.17cm}}l@{}c|S@{\hspace{-0.17cm}}l@{}c|S@{\hspace{-0.17cm}}l@{}c|S@{\hspace{-0.17cm}}l@{}c|}
\toprule
Parameter & Input & \multicolumn{3}{c}{No prior constraint} & \multicolumn{3}{c}{Orbit constrained} & \multicolumn{3}{c}{Mass \& orbit constrained} & \multicolumn{3}{c}{Mass constrained} \\
\midrule
$\log\,$$f_{\rm N_2}$          &   -0.11   &       -5.1  & $_{-3.47}^{+3.28}$ &  NC     &        -4.98  & $_{-3.50}^{+3.39}$ &  NC     &        -4.86  & $_{-3.52}^{+3.32}$ &  NC     &        -4.81  & $_{-3.37}^{+3.15}$ &  NC    \\
$\log\,$$f_{\rm O_2}$          &   -0.68   &      -0.77  & $_{-0.40}^{+0.35}$ &  C      &        -0.77  & $_{-0.39}^{+0.37}$ &  C      &        -0.82  & $_{-0.40}^{+0.37}$ &  C      &        -0.81  & $_{-0.36}^{+0.35}$ &  C     \\
$\log\,$$f_{\rm H_2O}$         &   -2.52   &      -2.57  & $_{-0.41}^{+0.38}$ &  C      &        -2.58  & $_{-0.40}^{+0.43}$ &  C      &        -2.62  & $_{-0.39}^{+0.38}$ &  C      &        -2.66  & $_{-0.35}^{+0.40}$ &  C     \\
$\log\,$$f_{\rm CO_2}$         &    -3.4   &      -6.53  & $_{-2.38}^{+2.47}$ &  NC     &        -6.28  & $_{-2.46}^{+2.51}$ &  NC     &         -6.2  & $_{-2.55}^{+2.46}$ &  NC     &        -6.54  & $_{-2.39}^{+2.68}$ &  NC    \\
$\log\,$$f_{\rm O_3}$          &   -6.15   &      -6.22  & $_{-0.25}^{+0.30}$ &  C      &        -6.18  & $_{-0.26}^{+0.33}$ &  C      &        -6.19  & $_{-0.17}^{+0.15}$ &  C      &        -6.35  & $_{-0.21}^{+0.28}$ &  C     \\
$\log\,$$f_{\rm CH_4}$         &    -5.7   &      -7.44  & $_{-1.74}^{+1.70}$ &  NC     &        -7.46  & $_{-1.70}^{+1.69}$ &  NC     &        -7.56  & $_{-1.65}^{+1.72}$ &  NC     &        -7.43  & $_{-1.79}^{+1.52}$ &  NC    \\
$\log\,$$p_{\rm surf}$               &    5.0    &       5.16  & $_{-0.41}^{+0.55}$ &  C      &         5.24  & $_{-0.40}^{+0.47}$ &  C      &         5.29  & $_{-0.39}^{+0.47}$ &  C      &         4.98  & $_{-0.40}^{+0.49}$ &  C     \\
$T$                       &   255.0   &     254.56  & $_{-27.08}^{+31.44}$ &  C      &       259.52  & $_{-26.66}^{+28.75}$ &  C      &       258.43  & $_{-26.53}^{+29.82}$ &  C      &       253.38  & $_{-29.94}^{+34.04}$ &  C     \\
$\log\,$$A_{\rm surf}$           &    -1.3   &      -1.28  & $_{-0.46}^{+0.38}$ &  NC     &        -1.39  & $_{-0.40}^{+0.41}$ &  NC     &        -1.35  & $_{-0.43}^{+0.35}$ &  NC     &        -1.17  & $_{-0.54}^{+0.40}$ &  NC    \\
$\log\,$$R_{\rm p}$           &    0.0    &       0.47  & $_{-0.47}^{+0.28}$ &  NC     &          0.0  & $_{-0.09}^{+0.12}$ &  C      &        -0.01  & $_{-0.07}^{+0.07}$ &  C      &         0.22  & $_{-0.29}^{+0.25}$ &  NC    \\
$\log\,$$M_{\rm p}$           &    0.0    &        0.8  & $_{-1.00}^{+0.75}$ &  NC     &         0.03  & $_{-0.62}^{+0.77}$ &  NC     &        -0.01  & $_{-0.05}^{+0.04}$ &  NA     &         -0.0  & $_{-0.05}^{+0.04}$ &  NA    \\
$\log\,$$\Delta p_{\rm c}$    &    4.0    &       2.33  & $_{-1.56}^{+1.74}$ &  NC     &         2.59  & $_{-1.76}^{+1.80}$ &  NC     &         2.69  & $_{-1.84}^{+1.66}$ &  NC     &         2.23  & $_{-1.59}^{+1.77}$ &  NC    \\
$\log\,$$p_{\rm c}$           &    4.78   &       4.81  & $_{-0.33}^{+0.34}$ &  C      &         4.88  & $_{-0.33}^{+0.31}$ &  C      &         4.87  & $_{-0.28}^{+0.27}$ &  C      &         4.66  & $_{-0.33}^{+0.33}$ &  C     \\
$\log\,$$\tau_{\rm c}$        &    1.0    &        1.0  & $_{-0.12}^{+0.23}$ &  C      &         1.01  & $_{-0.11}^{+0.19}$ &  C      &          1.0  & $_{-0.10}^{+0.17}$ &  C      &         0.96  & $_{-0.11}^{+0.21}$ &  C     \\
$\log\,$$f_{\rm c}$           &    -0.3   &      -0.26  & $_{-0.29}^{+0.19}$ &  C      &        -0.32  & $_{-0.32}^{+0.20}$ &  NC     &        -0.29  & $_{-0.20}^{+0.16}$ &  C      &        -0.19  & $_{-0.28}^{+0.14}$ &  C     \\
$a$                           &    1.0    &       3.68  & $_{-2.47}^{+3.63}$ &  NC     &          1.0  & $_{-0.10}^{+0.10}$ &  NA     &          1.0  & $_{-0.11}^{+0.10}$ &  NA     &         2.51  & $_{-1.62}^{+2.33}$ &  NC    \\
$\alpha$                      &    90.0   &      66.28  & $_{-45.75}^{+37.97}$ &  NC     &        90.14  & $_{-8.89}^{+8.61}$ &  NA     &        90.16  & $_{-8.55}^{+9.17}$ &  NA     &        56.84  & $_{-38.44}^{+40.83}$ &  NC    \\
\bottomrule
\end{tabular}
\begin{tablenotes}
\small
\item \textbf{Note.} For log-scale prior ranges, the parameters are Constrained (``C'') when the width of the 68\% confidence interval (corresponding to the interval between the mean value $-$1 and $+$1 standard deviation) is included within 1 log-unit (i.e., within one order of magnitude) for prior ranges spanning at least 5 log-units and within 0.5 log-units otherwise. For lin-scale prior ranges, parameters are Constrained (``C'') when the width of the 68\% confidence interval is smaller than 10\% of the full prior range width, and Tightly Constrained (``TC'') when smaller than $\pm 10\%$ of the input value. Parameters are Non Constrained (``NC'') otherwise, corresponding to either flat posterior distributions across the entire (or nearly) prior range or to posterior distributions with marked peak but also substantial tail towards extreme values. Detection strength is Not Applicable (``NA'') for parameters known \textit{a priori} (i.e., gaussian prior).
\end{tablenotes}
\end{threeparttable}
}
\end{table}

\begin{table}
\centering
\resizebox{1\textwidth}{!}{%
\begin{threeparttable}
\caption{Retrieval results comparison and associated level of constraint for the different cases of prior knowledge considered for vNIR and SNR = 15}
\label{tab:ret_results_vNIR_SNR15}
\begin{tabular}{lS|S@{\hspace{-0.17cm}}l@{}c|S@{\hspace{-0.17cm}}l@{}c|S@{\hspace{-0.17cm}}l@{}c|}
\toprule
Parameter & Input & \multicolumn{3}{c}{No prior constraint} & \multicolumn{3}{c}{Orbit constrained} & \multicolumn{3}{c}{Mass \& orbit constrained} \\
\midrule
$\log\,$$f_{\rm N_2}$          &   -0.11   &      -4.81  & $_{-3.35}^{+3.22}$ &  NC     &        -5.19  & $_{-3.32}^{+3.36}$ &  NC     &        -4.98  & $_{-3.38}^{+3.34}$ &  NC    \\
$\log\,$$f_{\rm O_2}$          &   -0.68   &      -0.76  & $_{-0.52}^{+0.41}$ &  C      &        -0.77  & $_{-0.53}^{+0.42}$ &  C      &        -0.88  & $_{-0.52}^{+0.44}$ &  C     \\
$\log\,$$f_{\rm H_2O}$         &   -2.52   &      -2.55  & $_{-0.45}^{+0.47}$ &  C      &        -2.52  & $_{-0.51}^{+0.46}$ &  C      &        -2.66  & $_{-0.47}^{+0.46}$ &  C     \\
$\log\,$$f_{\rm CO_2}$         &    -3.4   &      -5.83  & $_{-2.88}^{+2.38}$ &  NC     &        -6.27  & $_{-2.52}^{+2.59}$ &  NC     &        -6.19  & $_{-2.59}^{+2.54}$ &  NC    \\
$\log\,$$f_{\rm O_3}$          &   -6.15   &       -6.2  & $_{-0.30}^{+0.36}$ &  C      &        -6.16  & $_{-0.30}^{+0.35}$ &  C      &        -6.21  & $_{-0.21}^{+0.19}$ &  C     \\
$\log\,$$f_{\rm CH_4}$         &    -5.7   &      -7.24  & $_{-1.82}^{+1.70}$ &  NC     &        -7.29  & $_{-1.82}^{+1.71}$ &  NC     &        -7.48  & $_{-1.72}^{+1.74}$ &  NC    \\
$\log\,$$p_{\rm surf}$               &    5.0    &       5.17  & $_{-0.46}^{+0.56}$ &  NC     &         5.33  & $_{-0.49}^{+0.58}$ &  NC     &         5.34  & $_{-0.47}^{+0.57}$ &  NC    \\
$T$                       &   255.0   &     256.76  & $_{-34.37}^{+37.68}$ &  C      &       257.52  & $_{-36.10}^{+42.33}$ &  C      &       259.35  & $_{-35.89}^{+42.71}$ &  C     \\
$\log\,$$A_{\rm surf}$           &    -1.3   &      -1.22  & $_{-0.46}^{+0.44}$ &  NC     &        -1.38  & $_{-0.42}^{+0.42}$ &  NC     &        -1.28  & $_{-0.45}^{+0.35}$ &  NC    \\
$\log\,$$R_{\rm p}$           &    0.0    &       0.48  & $_{-0.45}^{+0.30}$ &  NC     &         0.01  & $_{-0.10}^{+0.13}$ &  C      &        -0.02  & $_{-0.08}^{+0.08}$ &  C     \\
$\log\,$$M_{\rm p}$           &    0.0    &       0.92  & $_{-1.05}^{+0.71}$ &  NC     &         0.13  & $_{-0.73}^{+0.90}$ &  NC     &         -0.0  & $_{-0.05}^{+0.04}$ &  NA    \\
$\log\,$$\Delta p_{\rm c}$    &    4.0    &       2.65  & $_{-1.74}^{+1.63}$ &  NC     &         2.63  & $_{-1.74}^{+1.70}$ &  NC     &         2.58  & $_{-1.74}^{+1.71}$ &  NC    \\
$\log\,$$p_{\rm c}$           &    4.78   &        4.8  & $_{-0.41}^{+0.41}$ &  C      &         4.89  & $_{-0.40}^{+0.40}$ &  C      &         4.89  & $_{-0.35}^{+0.35}$ &  C     \\
$\log\,$$\tau_{\rm c}$        &    1.0    &       1.01  & $_{-0.16}^{+0.33}$ &  C      &         1.02  & $_{-0.14}^{+0.24}$ &  C      &          1.0  & $_{-0.13}^{+0.21}$ &  C     \\
$\log\,$$f_{\rm c}$           &    -0.3   &      -0.31  & $_{-0.36}^{+0.21}$ &  NC     &        -0.33  & $_{-0.38}^{+0.23}$ &  NC     &        -0.28  & $_{-0.24}^{+0.18}$ &  C     \\
$a$                           &    1.0    &       3.68  & $_{-2.53}^{+3.71}$ &  NC     &          1.0  & $_{-0.10}^{+0.10}$ &  NA     &          1.0  & $_{-0.10}^{+0.10}$ &  NA    \\
$\alpha$                      &    90.0   &       63.0  & $_{-44.11}^{+42.09}$ &  NC     &        89.98  & $_{-9.01}^{+8.82}$ &  NA     &        89.66  & $_{-8.90}^{+8.51}$ &  NA    \\
\bottomrule
\end{tabular}
\end{threeparttable}
}
\end{table}

\begin{table}
\centering
\resizebox{1\textwidth}{!}{%
\begin{threeparttable}
\caption{Retrieval results comparison and associated level of constraint for the different cases of prior knowledge considered for vNIR and SNR = 10}
\label{tab:ret_results_vNIR_SNR10}
\begin{tabular}{lS|S@{\hspace{-0.17cm}}l@{}c|S@{\hspace{-0.17cm}}l@{}c|S@{\hspace{-0.17cm}}l@{}c|}
\toprule
Parameter & Input & \multicolumn{3}{c}{No prior constraint} & \multicolumn{3}{c}{Orbit constrained} & \multicolumn{3}{c}{Mass \& orbit constrained} \\
 \midrule 
$\log\,$$f_{\rm N_2}$          &   -0.11   &      -4.97  & $_{-3.50}^{+3.18}$ &  NC     &        -5.03  & $_{-3.36}^{+3.17}$ &  NC     &        -5.05  & $_{-3.35}^{+3.36}$ &  NC    \\
$\log\,$$f_{\rm O_2}$          &   -0.68   &      -1.02  & $_{-0.89}^{+0.59}$ &  NC     &        -0.98  & $_{-0.98}^{+0.61}$ &  NC     &        -1.07  & $_{-0.84}^{+0.63}$ &  NC    \\
$\log\,$$f_{\rm H_2O}$         &   -2.52   &       -2.6  & $_{-0.55}^{+0.65}$ &  NC     &        -2.51  & $_{-0.67}^{+0.63}$ &  NC     &        -2.68  & $_{-0.59}^{+0.63}$ &  NC    \\
$\log\,$$f_{\rm CO_2}$         &    -3.4   &       -6.1  & $_{-2.57}^{+2.90}$ &  NC     &        -6.14  & $_{-2.68}^{+2.88}$ &  NC     &         -6.0  & $_{-2.62}^{+2.58}$ &  NC    \\
$\log\,$$f_{\rm O_3}$          &   -6.15   &      -6.26  & $_{-0.31}^{+0.42}$ &  C      &         -6.1  & $_{-0.39}^{+0.41}$ &  C      &        -6.25  & $_{-0.26}^{+0.25}$ &  C     \\
$\log\,$$f_{\rm CH_4}$         &    -5.7   &      -7.42  & $_{-1.76}^{+1.82}$ &  NC     &        -7.27  & $_{-1.88}^{+1.91}$ &  NC     &        -7.37  & $_{-1.82}^{+1.77}$ &  NC    \\
$\log\,$$p_{\rm surf}$               &    5.0    &       5.28  & $_{-0.70}^{+0.70}$ &  NC     &         5.46  & $_{-0.68}^{+0.79}$ &  NC     &         5.37  & $_{-0.55}^{+0.61}$ &  NC    \\
$T$                       &   255.0   &     262.43  & $_{-55.46}^{+64.32}$ &  NC     &       264.67  & $_{-56.85}^{+68.99}$ &  NC     &       260.28  & $_{-53.81}^{+68.03}$ &  NC    \\
$\log\,$$A_{\rm surf}$           &    -1.3   &      -1.17  & $_{-0.55}^{+0.47}$ &  NC     &         -1.3  & $_{-0.45}^{+0.44}$ &  NC     &        -1.13  & $_{-0.55}^{+0.35}$ &  NC    \\
$\log\,$$R_{\rm p}$           &    0.0    &        0.5  & $_{-0.45}^{+0.26}$ &  NC     &         0.03  & $_{-0.12}^{+0.16}$ &  C      &        -0.02  & $_{-0.09}^{+0.09}$ &  C    \\
$\log\,$$M_{\rm p}$           &    0.0    &       0.95  & $_{-1.00}^{+0.71}$ &  NC     &         0.48  & $_{-0.87}^{+0.92}$ &  NC     &         -0.0  & $_{-0.05}^{+0.04}$ &  NA    \\
$\log\,$$\Delta p_{\rm c}$    &    4.0    &       2.55  & $_{-1.84}^{+1.93}$ &  NC     &         2.53  & $_{-1.76}^{+2.01}$ &  NC     &         2.78  & $_{-1.90}^{+1.81}$ &  NC    \\
$\log\,$$p_{\rm c}$           &    4.78   &       4.75  & $_{-0.58}^{+0.55}$ &  NC     &         4.93  & $_{-0.58}^{+0.57}$ &  NC     &         4.83  & $_{-0.49}^{+0.44}$ &  C     \\
$\log\,$$\tau_{\rm c}$        &    1.0    &       1.05  & $_{-0.22}^{+0.49}$ &  C      &         1.07  & $_{-0.22}^{+0.54}$ &  C      &         1.04  & $_{-0.20}^{+0.50}$ &  C     \\
$\log\,$$f_{\rm c}$           &    -0.3   &      -0.32  & $_{-0.40}^{+0.24}$ &  NC     &        -0.47  & $_{-0.47}^{+0.33}$ &  NC     &        -0.32  & $_{-0.37}^{+0.23}$ &  NC    \\
$a$                           &    1.0    &       3.84  & $_{-2.78}^{+3.46}$ &  NC     &         0.99  & $_{-0.10}^{+0.10}$ &  NA     &         0.99  & $_{-0.10}^{+0.10}$ &  NA    \\
$\alpha$                      &    90.0   &      67.34  & $_{-44.39}^{+42.43}$ &  NC     &        90.13  & $_{-9.15}^{+8.67}$ &  NA     &        89.08  & $_{-8.66}^{+8.59}$ &  NA    \\
\bottomrule
\end{tabular}
\end{threeparttable}
}
\end{table}

% ------------------------------------------------------------------------------------ %

\section{Applicability to other spectral coverages} \label{sec:appendix_spec_cover}
\begin{figure}
    \centering
    \includegraphics[width=0.8\textwidth, height=1.\textheight, keepaspectratio]{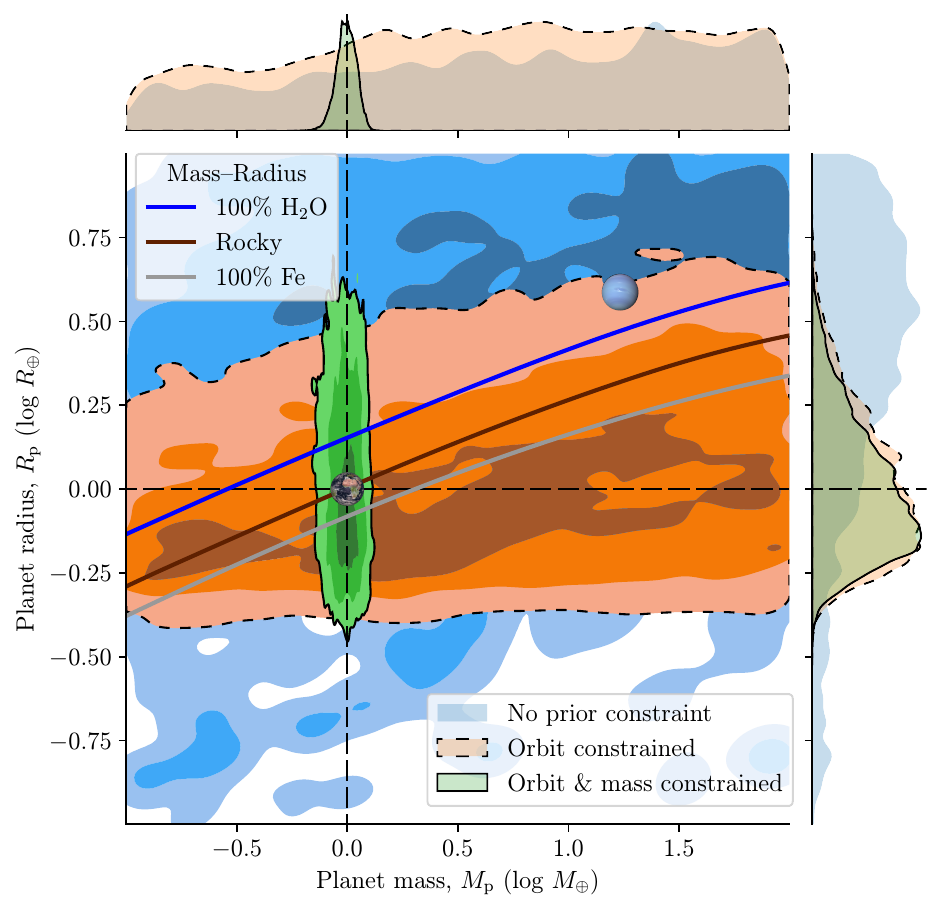}
    \caption{Mass--radius diagram (center) and marginal univariate posterior distributions (top and right side) for the different scenarios of prior knowledge (same as \autoref{fig:mass_radius}) considering a narrower spectral bandpass centered on ``red'' wavelengths ($\rm \lambda = [0.87,~1.05]~\upmu m$, $\rm res=140$; see \autoref{fig:fiducial_spectrum}) and $\rm SNR=10$.
    The contours of the 2D posterior distributions denote the 1, 2, and 3 sigma levels, encompassing 68\%, 95\%, and 99.7\% of the observed values, respectively. Earth-like input values are depicted with dashed horizontal and vertical lines. The small images of Earth and Neptune indicate their positions in the diagram. 
    For comparison, mass--radius relationships representative of different bulk compositions and interior structures are shown (pure \ce{H2O} assuming 1 mbar surface pressure level at 300 K, Earth-like rocky: 32.5\% \ce{Fe} + 67.5\% \ce{MgSiO3}, and pure iron; from \citealt{Zeng2019}).
    Retrieval results are given in \autoref{tab:ret_results_Red_SNR10}.
    }
    \label{fig:mass_radius_Red_SNR10}
\end{figure}

Atmospheric and surface properties inference and detection strengths are sensitive to the spectral region observed and to the SNR \citep[e.g.,][]{VonParis2013b, Feng2018, Damiano2020, Damiano2022, Konrad2022, Alei2022, Robinson2023, Damiano2023, Latouf2023a, Latouf2023b, Young2024a, Mettler2024}.
While we will specifically address the influence of the spectral coverage combined with the SNR on atmospheric retrievals of Earth analogs in another study, \autoref{fig:mass_radius_Red_SNR10} and \autoref{tab:ret_results_Red_SNR10} show that our conclusions are independent of the spectral coverage considered.
Indeed, for a narrower ``red'' bandpass ($\rm \lambda = [0.87,~1.05]~\upmu m$, with a spectral resolving power res = 140; see \autoref{fig:fiducial_spectrum}) and a SNR = 10 (specified at $\rm \lambda = 0.88~\upmu m$), a prior determination of the orbit-related parameters (i.e., orbital distance and phase angle) yields tight constraints on the planet radius. It is confidently constrained to be between 0.63 and 1.73 Earth radius ($\log R_{\rm p}/R_{\oplus} = -0.02^{+0.25}_{-0.19}$ for the 68\% confidence interval, compared to $\log R_{\rm p}/R_{\oplus} = 0.44^{+0.35}_{-0.51}$ without any prior knowledge).
On the other hand, the mass prior, when added to the orbit-related priors, does not offer substantial improvements on the planetary radius constraint ($\log R_{\rm p}/R_{\oplus} = -0.03^{+0.23}_{-0.16}$, corresponding to $R_{\rm p} = 0.93^{+0.67}_{-0.29}~ R_{\oplus}$; see \autoref{fig:mass_radius_Red_SNR10}), nor on the inference of other parameters (\autoref{tab:ret_results_Red_SNR10}).

\begin{table} 
\centering 
\resizebox{1.\textwidth}{!}{% 
\begin{threeparttable} 
\caption{Retrieval results comparison and associated level of constraint for the different cases of prior knowledge considered for the red bandpass (\autoref{fig:fiducial_spectrum}) and SNR = 10}
\label{tab:ret_results_Red_SNR10}
\begin{tabular}{lS|S@{\hspace{-0.17cm}}l@{}c|S@{\hspace{-0.17cm}}l@{}c|S@{\hspace{-0.17cm}}l@{}c|} 
\toprule 
Parameter & Input & \multicolumn{3}{c}{No prior constraint} & \multicolumn{3}{c}{Orbit constrained} & \multicolumn{3}{c}{Mass \& orbit constrained} \\ 
 \midrule
$\log\,$$f_{\rm N_2}$          &   -0.11   &      -4.54  & $_{-3.62}^{+3.34}$ &  NC     &        -4.98  & $_{-3.36}^{+3.29}$ &  NC     &         -5.1  & $_{-3.52}^{+3.37}$ &  NC    \\ 
$\log\,$$f_{\rm O_2}$          &   -0.68   &       -5.3  & $_{-3.41}^{+3.38}$ &  NC     &        -5.47  & $_{-3.10}^{+3.60}$ &  NC     &        -5.36  & $_{-3.18}^{+3.20}$ &  NC    \\ 
$\log\,$$f_{\rm H_2O}$         &   -2.52   &      -1.94  & $_{-1.59}^{+1.13}$ &  NC     &        -1.64  & $_{-1.23}^{+1.03}$ &  NC     &        -2.04  & $_{-1.15}^{+1.15}$ &  NC    \\ 
$\log\,$$f_{\rm CO_2}$         &    -3.4   &      -5.49  & $_{-3.26}^{+3.28}$ &  NC     &        -5.02  & $_{-3.37}^{+3.23}$ &  NC     &        -5.17  & $_{-3.23}^{+3.17}$ &  NC    \\ 
$\log\,$$f_{\rm O_3}$          &   -6.15   &      -6.13  & $_{-2.71}^{+2.92}$ &  NC     &        -6.04  & $_{-2.62}^{+2.64}$ &  NC     &        -5.88  & $_{-2.72}^{+2.60}$ &  NC    \\ 
$\log\,$$f_{\rm CH_4}$         &    -5.7   &      -6.83  & $_{-2.18}^{+2.42}$ &  NC     &        -6.44  & $_{-2.39}^{+2.34}$ &  NC     &        -6.66  & $_{-2.16}^{+2.39}$ &  NC    \\ 
$\log\,$$p_{\rm surf}$               &    5.0    &       4.88  & $_{-0.94}^{+1.44}$ &  NC     &         4.92  & $_{-0.75}^{+0.88}$ &  NC     &         4.73  & $_{-0.84}^{+0.82}$ &  NC    \\ 
$T$                       &   255.0   &     277.09  & $_{-66.23}^{+95.47}$ &  NC     &       286.48  & $_{-70.65}^{+92.67}$ &  NC     &       271.11  & $_{-67.25}^{+99.45}$ &  NC    \\ 
$\log\,$$A_{\rm surf}$           &    -1.3   &      -0.62  & $_{-0.73}^{+0.44}$ &  NC     &        -0.62  & $_{-0.67}^{+0.43}$ &  NC     &        -0.55  & $_{-0.60}^{+0.38}$ &  NC    \\ 
$\log\,$$R_{\rm p}$           &    0.0    &       0.44  & $_{-0.51}^{+0.35}$ &  NC     &        -0.02  & $_{-0.19}^{+0.25}$ &  C      &        -0.03  & $_{-0.16}^{+0.23}$ &  C     \\ 
$\log\,$$M_{\rm p}$           &    0.0    &       0.82  & $_{-1.06}^{+0.84}$ &  NC     &          0.7  & $_{-1.00}^{+0.88}$ &  NC     &         -0.0  & $_{-0.04}^{+0.04}$ &  NA    \\ 
$\log\,$$\Delta p_{\rm c}$    &    4.0    &       2.52  & $_{-1.68}^{+1.63}$ &  NC     &         2.29  & $_{-1.54}^{+1.66}$ &  NC     &         2.27  & $_{-1.56}^{+1.66}$ &  NC    \\ 
$\log\,$$p_{\rm c}$           &    4.78   &       2.79  & $_{-1.83}^{+1.65}$ &  NC     &         2.68  & $_{-1.83}^{+1.67}$ &  NC     &         2.69  & $_{-1.82}^{+1.55}$ &  NC    \\ 
$\log\,$$\tau_{\rm c}$        &    1.0    &      -0.07  & $_{-1.99}^{+1.95}$ &  NC     &         -0.0  & $_{-2.05}^{+1.96}$ &  NC     &        -0.25  & $_{-1.80}^{+2.05}$ &  NC    \\ 
$\log\,$$f_{\rm c}$           &    -0.3   &      -1.32  & $_{-1.19}^{+1.04}$ &  NC     &        -1.41  & $_{-1.05}^{+1.05}$ &  NC     &        -1.53  & $_{-0.99}^{+1.07}$ &  NC    \\ 
$a$                           &    1.0    &       3.34  & $_{-2.37}^{+3.83}$ &  NC     &          1.0  & $_{-0.10}^{+0.10}$ &  NA     &          1.0  & $_{-0.09}^{+0.10}$ &  NA    \\ 
$\alpha$                      &    90.0   &      65.79  & $_{-42.59}^{+47.86}$ &  NC     &        89.94  & $_{-9.60}^{+9.19}$ &  NA     &        90.14  & $_{-8.73}^{+8.33}$ &  NA    \\ 
\bottomrule
\end{tabular}
\end{threeparttable} 
} 
\end{table}

% ==================================================================================== %

\bibliography{biblio}
\bibliographystyle{aasjournal}

% ----------------------------------------------
% Display list of changes for revision
\listofchanges
% ----------------------------------------------

\end{document}